\documentclass[]{spie}  

\usepackage{amsmath,amsfonts,amssymb}
\usepackage{subcaption}
\usepackage{graphicx}
\usepackage{cite} 
\usepackage{times}
\usepackage{epsfig}
\usepackage{amsmath}
\usepackage{nccmath}
\usepackage{amssymb}
\usepackage{mwe}
\usepackage{acro}
\usepackage{amssymb}
\usepackage{xcolor,colortbl}
\usepackage{tabularx}
\usepackage{relsize}
\usepackage{pifont}
\usepackage{booktabs} 
\usepackage{multirow}
\usepackage{multicol}
\usepackage{adjustbox}
\usepackage{float}
\usepackage{graphicx}
\usepackage{makecell}
\usepackage{siunitx}
\usepackage{tabu}
\usepackage[colorlinks=true, allcolors=blue]{hyperref}
\usepackage[capitalize]{cleveref}

\title{Spatial Pathomics Toolkit for Quantitative Analysis of Podocyte Nuclei with Histology and Spatial Transcriptomics Data in Renal Pathology}

\author[a]{Jiayuan Chen*}
\author[b]{Yu Wang*}
\author[a]{Ruining Deng}
\author[a]{Quan Liu}
\author[a]{Can Cui}
\author[a]{Tianyuan Yao}
\author[a]{Yilin Liu}
\author[c,d]{Jianyong Zhong}
\author[c,d]{Agnes B. Fogo}
\author[c,d]{Haichun Yang}
\author[b]{Shilin Zhao}
\author[a,c,e]{Yuankai Huo}

\affil[a]{Department of Computer Science, Vanderbilt University, Nashville, TN, USA}
\affil[b]{Department of Biostatistics, Vanderbilt University Medical Center, Nashville, TN, USA}
\affil[c]{Department of Pathology, Microbiology and Immunology, Vanderbilt University Medical Center, Nashville, TN, USA}
\affil[d]{Department of Pathology, Microbiology and Immunology, Vanderbilt University Medical Center, Nashville, TN, USA}
\affil[d]{Department of Electrical and Computer Engineering, Vanderbilt University, Nashville, TN, USA}

\authorinfo{*Jiayuan Chen and Yu Wang contributed equally to this paper\\Corresponding author: Shilin Zhao: E-mail: shilin.zhao.1@vumc.org; Yuankai Huo: E-mail: yuankai.huo@vanderbilt.edu}

\begin{document} 
\maketitle

\begin{abstract}
Podocytes, specialized epithelial cells that envelop the glomerular capillaries, play a pivotal role in maintaining renal health. The current description and quantification of features on pathology slides are limited, prompting the need for innovative solutions to comprehensively assess diverse phenotypic attributes within Whole Slide Images (WSIs). In particular, understanding the morphological characteristics of podocytes, terminally differentiated glomerular epithelial cells, is crucial for studying glomerular injury. This paper introduces the Spatial Pathomics Toolkit (SPT) and applies it to podocyte pathomics. The SPT consists of three main components: (1) instance object segmentation, enabling precise identification of podocyte nuclei; (2) pathomics feature generation, extracting a comprehensive array of quantitative features from the identified nuclei; and (3) robust statistical analyses, facilitating a comprehensive exploration of spatial relationships between morphological and spatial transcriptomics features. The SPT successfully extracted and analyzed morphological and textural features from podocyte nuclei, revealing a multitude of podocyte morphomic features through statistical analysis. Additionally, we demonstrated the SPT's ability to unravel spatial information inherent to podocyte distribution, shedding light on spatial patterns associated with glomerular injury. By disseminating the SPT, our goal is to provide the research community with a powerful and user-friendly resource that advances cellular spatial pathomics in renal pathology. The toolkit’s implementation and its complete source code are made openly accessible at the GitHub repository: \url{https://github.com/hrlblab/spatial_pathomics}.

\end{abstract}



  


 \begin{figure*}[h]
\begin{center}
\includegraphics[width=1\linewidth]{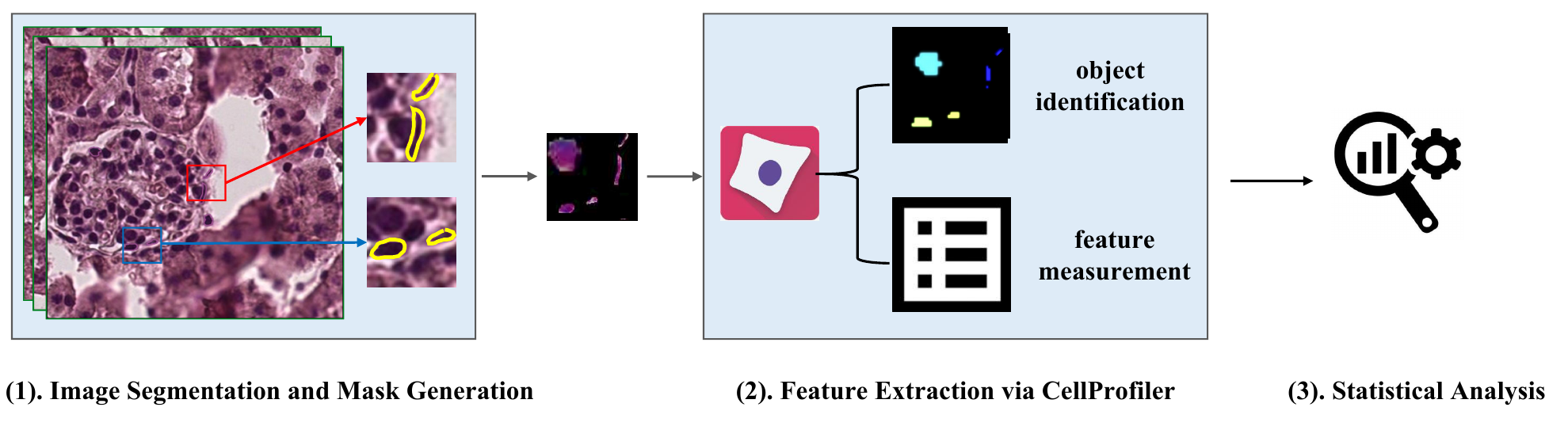}
\end{center}
\caption{ Overview of the experiment: segmentation and mask generation of podocytes from pathological images, extraction of features and statistical analysis.}
\label{fig:exp1}
\end{figure*}

\section{INTRODUCTION}
\label{sec:intro}  
The emerging field of Pathomics leverages image analysis to quantify diverse phenotypic attributes across cells, nuclei, and elements in Whole Slide Images (WSIs), promising a deeper understanding of pathological variations~\cite{gupta2019emergence,holscher2023next,fogo2023learning}. WSIs, with their complex histological landscapes containing numerous objects, drive research efforts to identify and quantify histological features, enabling exploration of biological behavior across various cancer phenotypes. Digital pathology facilitates quantitative assessment of diagnostic features, providing crucial data on cell types, tissue structures, tumor measurements, invasive borders, growth patterns, nuclear characteristics, and necrosis percentages throughout WSIs. Pathomics applications encompass nucleus morphology, cell type classification, spatial tumor characterization, and biomarker-labeled cell estimation through immunohistochemistry.

In contrast to tumors, renal pathology is inherently more complex due to the presence of over 20 types of cells within the kidney. Podocytes, as terminally differentiated epithelial cells, play a pivotal role in glomerular function, making their comprehensive analysis essential for understanding renal health and diseases. But their current understanding, mainly based on cell counts, lacks insight into intricate functions. The emerging field of Pathomics employs image analysis to quantify diverse phenotypic attributes across cells, nuclei, and elements in WSIs, promising a deeper grasp of phenotypic variations in pathological contexts. 
To bridge the existing gap and enable a comprehensive analysis of morphological and textural features from podocyte nuclei, we present the Spatial Pathomics Toolkit (SPT). 

The SPT serves as a versatile resource specifically designed for the extraction and analysis of quantitative features from podocyte nuclei in renal pathology. Comprising three essential components, SPT enables an in-depth investigation of podocyte nuclei. Firstly, it requires instance object segmentation, ensuring the precise identification and delineation of individual podocyte nuclei. This accurate segmentation allows SPT to focus on specific nuclei and explore their unique characteristics. This accurate segmentation allows researchers to focus on specific nuclei and explore their unique characteristics. Secondly, SPT provides pathomics feature generation, enabling the extraction of an extensive array of quantitative features from the identified podocyte nuclei. This feature extraction process captures crucial morphological and textural information, facilitating a comprehensive understanding of podocyte characteristics. Lastly, SPT empowers robust statistical analyses, enabling researchers to delve into spatial relationships between morphological and spatial transcriptomics features. Leveraging advanced statistical methods, researchers can gain valuable insights into the spatial distribution and organization of these nuclei within renal pathology samples.

In this study, we practically apply the SPT to comprehensively investigate podocyte nuclei within both diseased and normal kidney WSIs. By demonstrating the effectiveness of SPT, we uncover essential spatial information inherent in the distribution of podocyte nuclei. Our objective is to disseminate SPT as a powerful and user-friendly resource, empowering the research community to advance their understanding of podocyte-related renal pathologies and explore broader applications in life sciences and computational research. The complete implementation of the toolkit, along with its source code, is made openly accessible on the GitHub repository: \url{https://github.com/hrlblab/spatial_pathomics}.

\begin{table}[h]
    \centering
    \begin{tabular}{c|ccccc}
    \toprule
     Feature    & Area\&Shape & Texture & Intensity & IntensityDistribution & All\\
     \midrule
    Dimension  &  108 & 52 & 25 & 72 & 257\\
    \bottomrule
    \end{tabular}
    \caption{Feature dimension distribution.}
    \label{tab:1}
\end{table}

\begin{table}[h]
    \centering
\begin{tabular}{l|l}
\toprule
{Feature Type} &  {Feature Information} \\
\midrule
\multirow{2}{*}{Size \& Shape} & Area, Perimeter, Eccentricity, Solidity,\\
&MajorAxisLength, MinorAxisLength, ... \\
\midrule
\multirow{2}{*}{Texture} & Correlation, IDM (Inverse Difference Moment),  \\
&ASM (Angular Second Moment), Entropy, ... \\
\midrule
\multirow{2}{*}{Intensity} & Mean Intensity, Integrated Intensity, Total Intensity,  \\
&Max Intensity, Min Intensity, Median Intensity, ... \\
\midrule
\multirow{2}{*}{Intensity Distribution} & Intensity Mean, Intensity Median, Intensity StdDev,  \\
&Intensity Range, Intensity Mode, Intensity CV, ... \\
\bottomrule
\end{tabular}
\caption{Feature information partial summary for Size \& Shape, Texture, Intensity, and Intensity Distribution.}
   \label{tab:2}
\end{table}

\section{Method}

\subsection{Image Segmentation and Mask Generation}

To segment glomeruli-containing images, a total of 458 images (512$\times$512) were selected from 4 WSIs. For segmenting podocytes in mouse glomerular images, an efficient option is to utilize
Qupath \cite{Bankhead099796}, a digital pathology platform renowned for its user-friendly image annotation interface. With Qupath, manual annotation of specific cell nuclei was performed by an experienced pathologist, ensuring precise and accurate segmentation. 
Following the annotation of nucleus regions using either method, pixels outside the designated areas were masked to facilitate subsequent analysis focused on podocytes.

\subsection{Feature Extraction via CellProfiler}
Post-annotation and masking, the segmented cell nuclei were subjected to feature extraction. For this purpose, the CellProfiler Analyst software package \cite{cellprofiler}  was chosen due to its capacity for extracting quantitative and morphological features of cells and cellular structures from microscopy images. The feature extraction workflow encompassed two primary tasks: object identification and feature measurement.

Object identification was achieved by employing a thresholding algorithm to segment and label the nuclei in the masked images, resulting in distinct objects corresponding to individual cell nuclei. Once the objects were successfully identified, subsequent feature measurements were performed by analyzing the pixels within each object. The extracted features encompassed essential characteristics such as size, shape, textural patterns, and intensity of the labeled nuclear structures. Specific features included area and perimeter to determine nucleus size and boundary, major and minor axis lengths to assess nucleus elongation or circularity, and texture features derived from gray-level co-occurrence matrices, such as angular second moment, contrast, correlation, inverse difference moment, and entropy, to describe texture or pattern variations within the nuclei. Additionally, intensity features, including mean, median, maximum, and minimum intensity values, were quantified to assess brightness or fluorescence expression within each nucleus. Subsequently, the calculated feature values were exported for further downstream analysis, facilitating a deeper understanding of cell characteristics, pattern recognition, and correlation identification, ultimately advancing knowledge of cellular behavior and biological processes.

\subsection{Statistical Analysis}
Data analysis was conducted using R version 4.2.1 \url{https://www.r-project.org/}. The extracted morphomic features were processed based on their distribution patterns. Log transformation was applied to highly skewed data. For data with negative values or persistent skewness post-log transformation, Box Cox transformation was utilized. After transformations, data were manually inspected and scaled in preparation for subsequent statistical analyses. Principal component analysis (PCA) was employed for dimensionality reduction. Features with an absolute loading greater than 0.15 in any principal component were considered contributory to that PC. In comparative analyses, t-test was used to compare principal components, and Wilcoxon rank sum test was used to compare features between disease and normal podocytes. Spearman correlation was used to assess relationships between features and spatial transcriptomics gene expression. Features with an absolute correlation coefficient greater than 0.9 were considered as highly correlated. Hierarchical clustering was used in heatmaps. The circos plot was generated with R package circlize~\cite{circlize}. Spatial related data analysis was executed with R package Seurat~\cite{Seurat}. Multiple comparison adjustment was performed by False Discovery Rate (FDR) when needed. The level of statistical significance was set at p \textless 0.05 or FDR \textless 0.05 for all analyses.

\section{Data and Experiments}
\subsection{Data}
The data set comprised transgenic mice expressing the human diphtheria toxin receptor (DTR) in segments S1 and S2 of the proximal tubules to induce acute tubular injury~\cite{lim2017tubulointerstitial}. Diphtheria toxin (DT) was administered to the DTR+ mice, while the wild type (WT) mice received DT as a normal control. All mice were sacrificed six weeks after DT injection. Renal function and kidney morphology analysis revealed significant tubular injury but minor glomerular changes in DTR+ mice. The harvested kidneys were fixed in paraformaldehyde, and 10\textmu m sections were cut and subsequently stained with hematoxylin and eosin (H\&E). For the investigation, we employed 2 WSIs from DTR+ mice and 2 WSIs from WT mice, all of which were scanned under a 40$\times$ magnification, as well as 10$\times$ Visium
spatial transcriptomics acquisition. We extracted the patches with 55 \textmu m diameter.

\begin{figure*}[h]
\begin{center}
\includegraphics[width=1\linewidth]{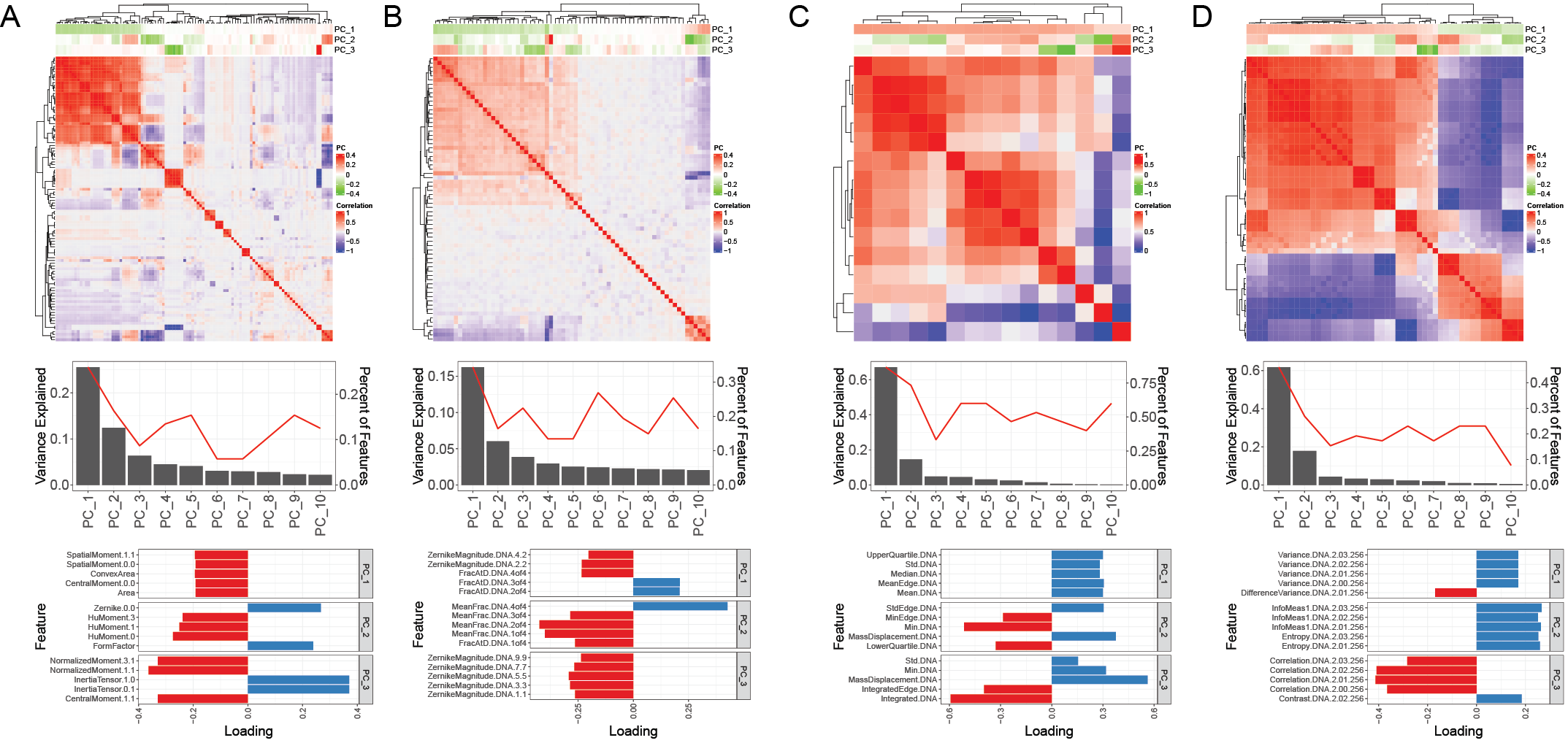}
\end{center}
\caption{Morphomic feature characteristics. (A) AreaShape features; Top Panel: Heatmap visualization of clustered correlation coefficients between features. Middle Panel: Percentage variance explained by the top 10 PCs (bar graph) and corresponding number of features contributed (line graph). Bottom Panel: Highlight of top 5 features with the most significant contribution to PC1-PC3. (B) Radial distribution features; (C) Intensity features; (D) Texture features; }
\label{fig:exp}
\end{figure*}

\begin{figure*}[h]
\begin{center}
\includegraphics[scale=1]{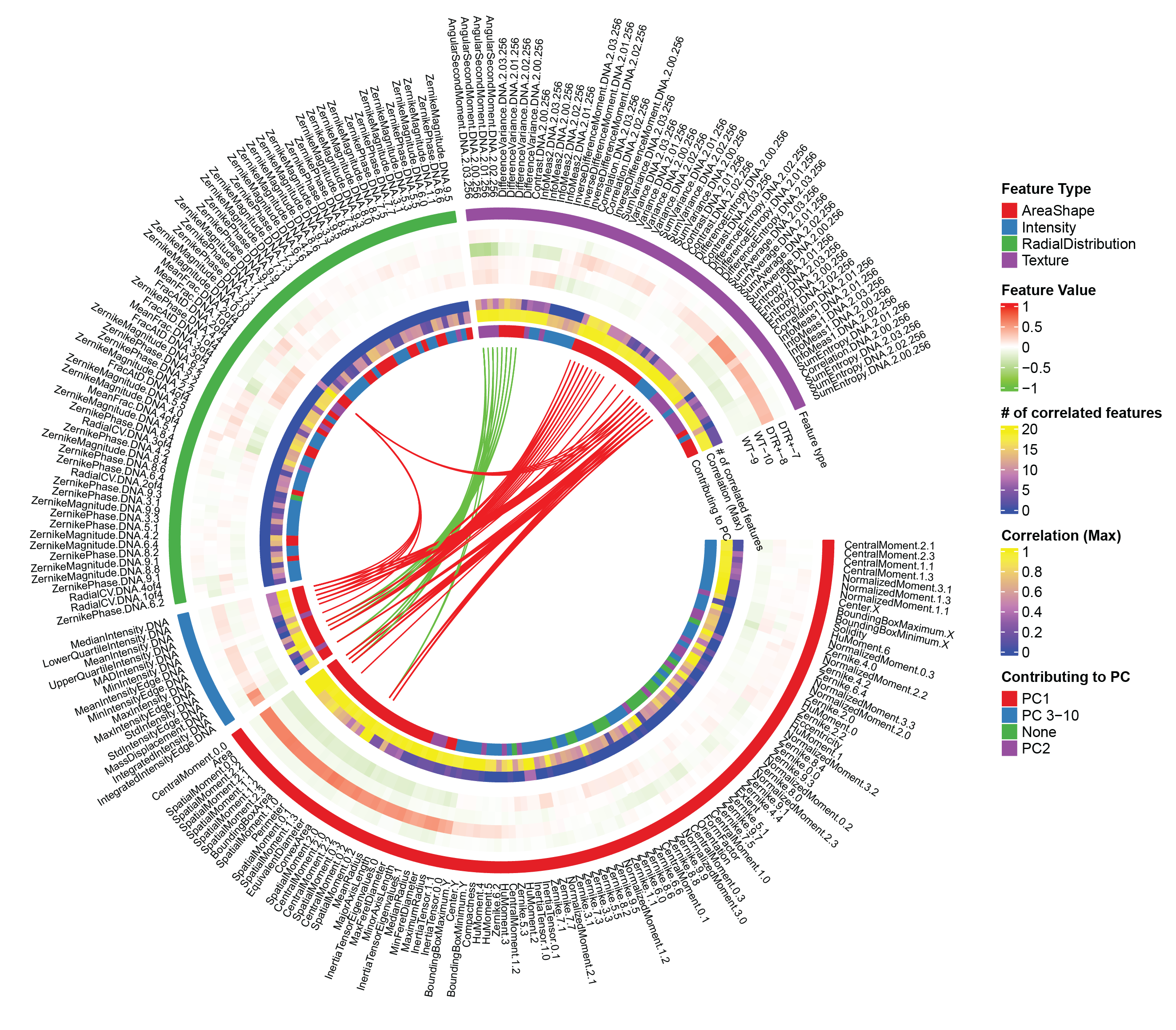}
\end{center}
\caption{Visualization of Morphomic Features Using a Circos Plot. Outermost Layer (1): Type of Feature; Layer (2): Average values of features across four samples; Layer (3): Count of features with strong correlations; Layer (4): Highest correlation coefficient with other features; Layer (5): Principal component the feature contributes most to; Innermost Layer (6): Lines indicating pairs of highly correlated features.}
\label{fig:exp}
\end{figure*}

\begin{figure*}[h]
\begin{center}
\includegraphics[width=0.8\linewidth]{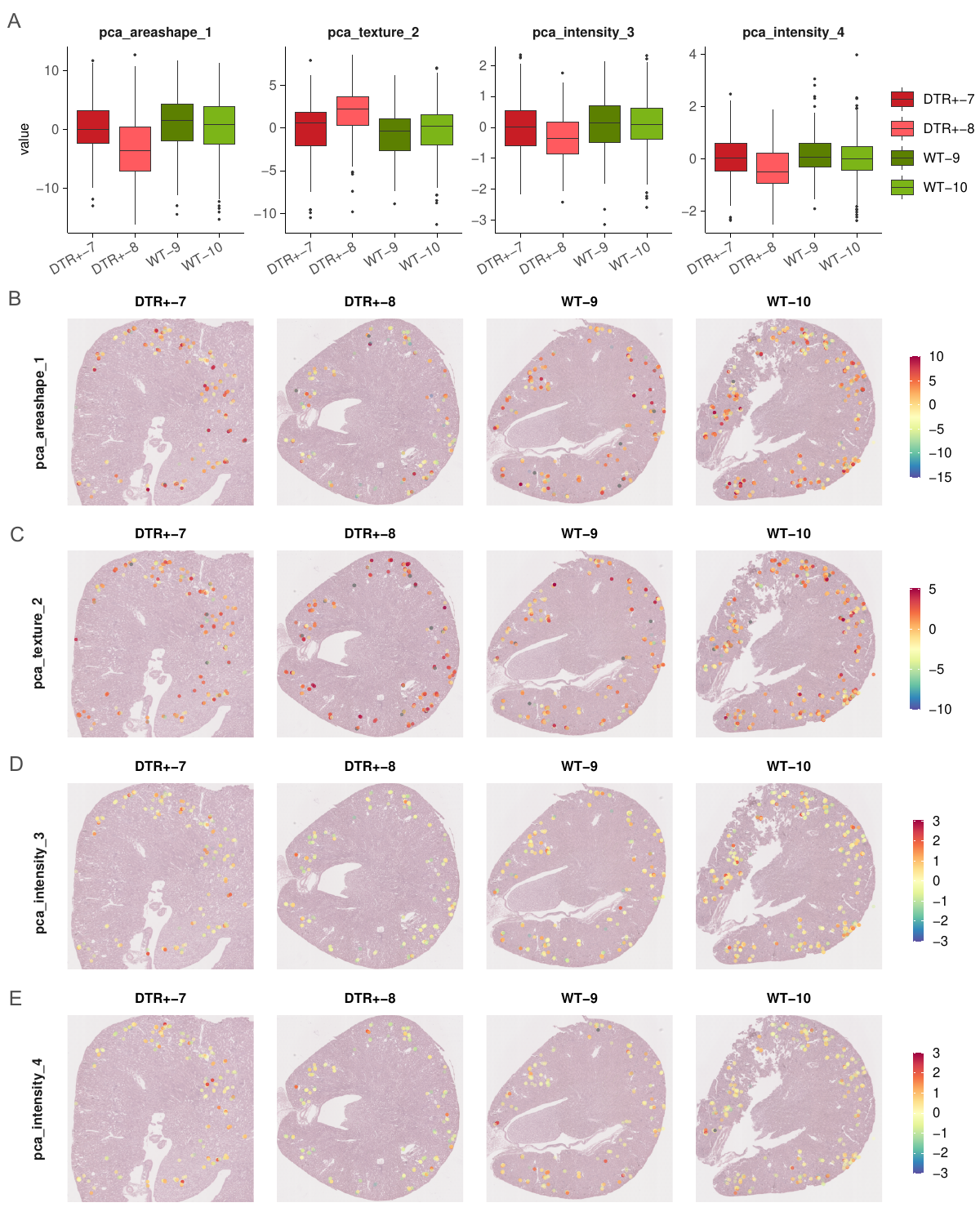}
\end{center}
\caption{Differential morphomic feature PCs between disease and normal podocytes. (A) Boxplot for the four significantly different morphomic feature values among four samples; (B) Spatial plot for AreaShape PC1; (C) Spatial plot for Texture PC2; (D) Spatial plot for Intensity PC3; (E) Spatial plot for Intensity PC4;}
\label{fig:exp}
\end{figure*}

\subsection{Experiments}
Figure \ref{fig:exp1} shows the overview of the experiment. During the experiment, we batch import images into QuPath for annotation. The regions of pathological podocytes in each image are exported to a geojson file for subsequent masking. The masked images are then fed into CellProfiler for feature extraction. Each dimension of the feature describes an aspect of biological phenotypic information. During the process of feature extraction, we begin by utilizing a thresholding technique to identify the region of the nucleus in the image, which determines the pixels of the target nucleus through an adaptive threshold. Then we invoke the analysis module of CellProfiler to extract features about shape, size, texture, and intensity. The dimensions of each category of features are shown in Table \ref{tab:1}. We list part of the feature information in Table \ref{tab:2}. Using a single Intel Core i5 CPU, the process of feature extraction on 458 images required 0.5 hour.

\section{Results}
\subsection{Overview of Morphological features}
We began morphological features analysis by examining the correlation of different features, subsequently conducting PCA for each feature type. The Area features demonstrated a dominant, highly correlated cluster, complemented by smaller clusters and isolated features (Figure 2A, top panel). The principal components revealed that the first component explained more than 20\% variance in the data, with contributions primarily from 20+ features, especially features within the dominant cluster. The second component explained more than 10\% of variance and were contributed by 16 features from four small clusters. Other PCs explained less than 5\% of variance and were contributed by 5-10 features respectively (Figure 2A, top and middle panel). Key contributors to PC1 included two Spatial Moment features, a Central Moment feature, Area, and Convex Area. For PC2, three HuMoments features, a Zernike feature, and FormFactor were predominant (Figure 2A, Bottom panel). 

Similarly, the Radial distribution features exhibited several correlated clusters alongside unique features (Figure 2B, top and middle panel). The first component from PCA explained more than 15\% of variance and was shaped by 20+ features predominantly from two primary correlation clusters. FracAtD and ZernikeMagnitude features were pivotal for PC1, while MeanFrac and another FracAtD feature dominated PC2 contributions (Figure 2B, Bottom panel).

In contrast, Intensity and Texture features were largely characterized by a single, heavily correlated cluster with few numbers of independent features (Figure 2C, 2D, top panel). The first component of Intensity or Texture accounted for an impressive over 60\% variance, hinting at a high similarity and redundancy among these types of features (Figure 2C, 2D), which was also consistent with the dominate correlation clusters in the correlation heatmap. The second component explained about 14\% (Intensity) or 17\% (texture) of variance. Each of the rest of components only explained less than 5\% variance. The prominent Intensity features for its first component were UpperQuantile, Std, Median, MeanEdge, and Mean, whereas the Texture domain was led by Variance-related attributes. 

In summary, the result highlights redundancy in all feature types but underscores distinct information in Area and Radial features compared to the homogeneity in Intensity and Texture.

\subsection{Morphological characteristics of podocytes}
We used circos plot to integrate and visualize the distribution of morphomic features across four samples as well as feature characteristics and annotations. The features were color-coded by type in the plot's outer layer, while a subsequent heatmap (after hierarchical clustering) illustrated the average feature values across samples. Notably, DTR+ samples exhibited distinct patterns compared to WT samples, with DTR+-8 standing out in terms of elevated AreaShape, Texture, and Intensity values.

Our correlation analysis of morphomic features revealed that most of them only correlated with limited number of other features (indicated by blue in the 6th layer, \# of correlated features) except for features in the large cluster in AreaShape (bottom left, yellow). The 7th layer's color bar (Correlation (Max)), showcasing the maximal correlation coefficient, underscored that most features had strong associations with at least one other (yellow hues denoting correlations exceeding 0.8). Yet, a substantial portion of RadialDistribution and half of Area features remained distinct with minimal correlations (depicted in blue or purple, suggesting correlations below 0.2). The inner layer lines pinpointed highly correlated feature pairs across types, with a pronounced correlation between Intensity and Texture. RadialDistribution, however, remained uniquely distinct with minimal correlation pairs. Lastly, an analysis on principal components highlighted varying feature contributions: while AreaShape and Texture had a mixed contribution across PCs, Intensity leaned heavily on the first component, and RadialDistribution primarily on PC1 and PC3-10.

\subsection{Differential morphomic features between disease and normal podocytes}
Due to the redundancy observed in morphomic features, we initially employed a t-test to assess the distribution differences of the first five principal components (PCs) across samples in both groups (Figure 4A). Out of the 20 PCs evaluated, four showed significant difference between the groups: AreaShape PC1, Texture PC2, and Intensity PC3 and PC4 (Figure 4A).

Subsequently, these PCs were projected back onto histology image patches, visualizing their values and spatial distribution among samples. Consistent with the boxplot findings, both AreaShape PC1 (Figure 4B) and Texture PC2 (Figure 4C) manifested marked differences in values between the two groups when observed in the Spatial value plot. Notably, we also noticed that neighboring podocytes often have similar values, suggesting that podocytes from the same glomerulus have similar morphological characteristics. 

To further assess the spatial autocorrelation between PC values and their locations, we computed the Moran's I classic statistic. Our results corroborated the initial observations: both AreaShape PC1 (p=5.13e-05) and Texture PC2 (p=4.09e-04), along with Intensity PC3 (p=0.0017), showcased significant correlations with their respective spatial patterns.


\subsection{Gene-Linked spatial patterns of differential morphomic features}
After exploring the pathomics features relationship, we further identify the differential features between DTR+ and WT podocytes. A total of 58 features displayed differential values between these groups with an FDR \textless 0.05 (Fig. 5A; AreaShape = 32, Intensity = 3, RadialDistribution = 6, Texture = 17). Intriguingly, the majority of feature values were elevated in the DTR+ group compared to the WT (Fig. 5B).

To better understand these differential features, we integrated the morphomic features with the spatial transcriptomics data from corresponding samples, and subsequently assessed the correlation between each differential feature and the genes highly expressed in individual samples (Fig. 5C, E, G, I). A notable observation was the correlation of several differential features, specifically within the AreaShape category, with the expression of certain genes, but only within the DTR+ group (Fig. 5D, F). Specifically, morphomic feature AreaShape.InertiaTensorEigenvalues.0, AreaShape.MajorAxisLength, and AreaShape.MaxFeretDiameter were positively correlated with Eci2 gene in Sample DTR+-7 and negatively correlated with Pnn gene in Sample DTR+-8 (Fig. 5C, E). Eci2 gene encodes a member of the hydratase/isomerase superfamily and is a key mitochondrial enzyme involved in beta-oxidation of unsaturated fatty acids \cite{RN35}. PNN, (Pinin, Desmosome Associated Protein) is a multiple functional protein, and plays roles in embryonic development, cellular differentiation, tumorigenesis, and metastasis \cite{RN37}. PNN was reported to function as an oncogenic factor by reducing apoptosis and promoting cell migration and invasion in renal cell carcinoma \cite{RN36}. Additionally, AreaShape.InertiaTensor.0.0 also illustrated opposite correlation patterns with different genes across the two DTR+ samples (Fig. 5H, J). For instance, Rhbdf1, known for its growth factor binding and endopeptidase activities, when suppressed, can inhibit tumor growth, including in the kidney \cite{RN38}. Similarly, SUGCT's primary role involves the conversion of glutaric acid into glutaryl-CoA \cite{RN40,RN41}. Aberrations in SUGCT have been linked with Glutaric Aciduria III and Trichothiodystrophy \cite{RN39}.

Collectively, these insights shed light on potential molecular mechanisms underpinning the observed pathomics features.

\section{Conclusion}
The SPT toolkit represents an open-source tool for spatial pathomics analyses, linking histology pathomics, and spatial transcriptomics. SPT identifies spatial feature patterns associated with glomerular injury and correlates feature distribution with spatial gene expression. By making the SPT available to the broader research community, we expect to empower pathologists with a new tool for quantitative histopathology and spatial biology analysis, paving the way for improving the understanding of kidney disease.

\section{ACKNOWLEDGMENTS}       
This work has not been submitted for publication or presentation elsewhere. This work is supported in part by NIH R01DK135597(Huo), DoD HT9425-23-1-0003(HCY), and NIH NIDDK DK56942(ABF).

\newpage
\begin{figure*}[ht!]
\begin{center}
\includegraphics[width=0.7\linewidth]{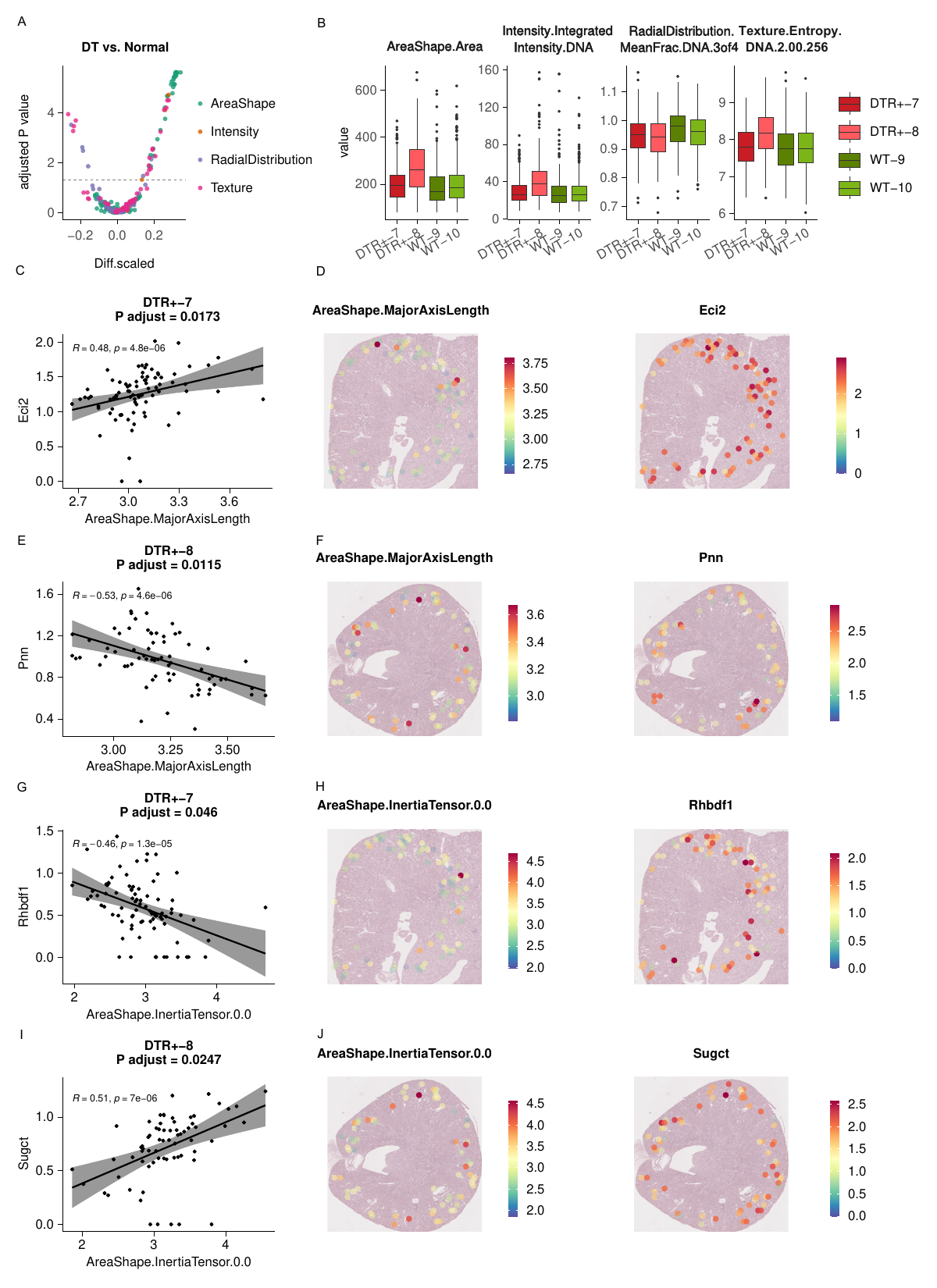}
\end{center}
\caption{Spatial patterns of differential pathomics features between DTR+ and WT groups and the correlation with gene expression in DTR+ samples. A. Violin plot of the differential pathomics features between DTR+ and WT. Colors show the four main feature categories. X-axis is the scaled mean differences, while y-axis shows the negative log10 transformed adjusted P values. B. The boxplots show the values of the top selected features in each group across 4 samples. C, E, G, I, the scatter plots show the correlation of one of the differential features in A and the gene expression in corresponding sample. Correlation rhos, raw p values and adjusted p values are shown in the plots. D, F, H, J, the spatial expression patterns of the feature-gene pairs. Feature values were summarized to spot level to match the transcriptomics data by calculating the mean of the features if multiple cells fall into one spot. Normalized expression values were shown for gene expression.}
\label{fig:exp}
\end{figure*}

\bibliography{main} 

\begin{thebibliography}{10}

\bibitem{gupta2019emergence}
Gupta, R., Kurc, T., Sharma, A., Almeida, J.~S., and Saltz, J., ``The emergence
  of pathomics,'' {\em Current Pathobiology Reports}~{\bf 7},  73--84 (2019).

\bibitem{holscher2023next}
H{\"o}lscher, D.~L., Bouteldja, N., Joodaki, M., Russo, M.~L., Lan, Y.-C.,
  Sadr, A.~V., Cheng, M., Tesar, V., Stillfried, S.~V., Klinkhammer, B.~M.,
  et~al., ``Next-generation morphometry for pathomics-data mining in
  histopathology,'' {\em Nature Communications}~{\bf 14}(1),  470 (2023).

\bibitem{fogo2023learning}
Fogo, A.~B., ``Learning from deep learning and pathomics,'' {\em Kidney
  International}  (2023).

\bibitem{Bankhead099796}
Bankhead, P., Loughrey, M.~B., Fern{\'a}ndez, J.~A., Dombrowski, Y., McArt,
  D.~G., Dunne, P.~D., McQuaid, S., Gray, R.~T., Murray, L.~J., Coleman, H.~G.,
  James, J.~A., Salto-Tellez, M., and Hamilton, P.~W., ``Qupath: Open source
  software for digital pathology image analysis,'' {\em Scientific Reports}
  (2017).

\bibitem{cellprofiler}
Carpenter, A.~E., Jones, T.~R., Lamprecht, M.~R., Clarke, C., Kang, I.~H.,
  Friman, O., Guertin, D.~A., Chang, J.~H., Lindquist, R.~A., Moffat, J.,
  et~al., ``Cellprofiler: image analysis software for identifying and
  quantifying cell phenotypes,'' {\em Genome biology}~{\bf 7},  1--11 (2006).

\bibitem{circlize}
Gu, Z., Gu, L., Eils, R., Schlesner, M., and Brors, B., ``circlize implements
  and enhances circular visualization in r,'' {\em Bioinformatics}~{\bf
  30}(19),  2811--2812 (2014).

\bibitem{Seurat}
Satija, R., Farrell, J.~A., Gennert, D., Schier, A.~F., and Regev, A.,
  ``Spatial reconstruction of single-cell gene expression data,'' {\em Nat
  Biotechnol}~{\bf 33}(5),  495--502 (2015).
\newblock Satija, Rahul Farrell, Jeffrey A Gennert, David Schier, Alexander F
  Regev, Aviv eng.

\bibitem{lim2017tubulointerstitial}
Lim, B.~J., Yang, J.~W., Zou, J., Zhong, J., Matsusaka, T., Pastan, I., Zhang,
  M.-Z., Harris, R.~C., Yang, H.-C., and Fogo, A.~B., ``Tubulointerstitial
  fibrosis can sensitize the kidney to subsequent glomerular injury,'' {\em
  Kidney international}~{\bf 92}(6),  1395--1403 (2017).

\bibitem{RN35}
van Weeghel, M., te~Brinke, H., van Lenthe, H., Kulik, W., Minkler, P.~E.,
  Stoll, M.~S., Sass, J.~O., Janssen, U., Stoffel, W., Schwab, K.~O., Wanders,
  R.~J., Hoppel, C.~L., and Houten, S.~M., ``Functional redundancy of
  mitochondrial enoyl-coa isomerases in the oxidation of unsaturated fatty
  acids,'' {\em FASEB J}~{\bf 26}(10),  4316--26 (2012).

\bibitem{RN37}
Leu, S., ``The role and regulation of pnn in proliferative and non-dividing
  cells: Form embryogenesis to pathogenesis,'' {\em Biochem Pharmacol}~{\bf
  192},  114672 (2021).

\bibitem{RN36}
Jin, M., Li, D., Liu, W., Wang, P., Xiang, Z., and Liu, K., ``Pinin acts as a
  poor prognostic indicator for renal cell carcinoma by reducing apoptosis and
  promoting cell migration and invasion,'' {\em J Cell Mol Med}~{\bf 25}(9),
  4340--4348 (2021).

\bibitem{RN38}
Yan, Z., Zou, H., Tian, F., Grandis, J.~R., Mixson, A.~J., Lu, P.~Y., and Li,
  L.~Y., ``Human rhomboid family-1 gene silencing causes apoptosis or autophagy
  to epithelial cancer cells and inhibits xenograft tumor growth,'' {\em Mol
  Cancer Ther}~{\bf 7}(6),  1355--64 (2008).

\bibitem{RN40}
Niska-Blakie, J., Gopinathan, L., Low, K.~N., Kien, Y.~L., Goh, C. M.~F.,
  Caldez, M.~J., Pfeiffenberger, E., Jones, O.~S., Ong, C.~B., Kurochkin,
  I.~V., Coppola, V., Tessarollo, L., Choi, H., Kanagasundaram, Y., Eisenhaber,
  F., Maurer-Stroh, S., and Kaldis, P., ``Knockout of the non-essential gene
  sugct creates diet-linked, age-related microbiome disbalance with a
  diabetes-like metabolic syndrome phenotype,'' {\em Cell Mol Life Sci}~{\bf
  77}(17),  3423--3439 (2020).

\bibitem{RN41}
Marlaire, S., Van~Schaftingen, E., and Veiga-da Cunha, M., ``C7orf10 encodes
  succinate-hydroxymethylglutarate coa-transferase, the enzyme that converts
  glutarate to glutaryl-coa,'' {\em J Inherit Metab Dis}~{\bf 37}(1),  13--9
  (2014).

\bibitem{RN39}
Leandro, J., Bender, A., Dodatko, T., Argmann, C., Yu, C., and Houten, S.~M.,
  ``Glutaric aciduria type 3 is a naturally occurring biochemical trait in
  inbred mice of 129 substrains,'' {\em Mol Genet Metab}~{\bf 132}(2),
  139--145 (2021).

\end{thebibliography}
\bibliographystyle{spiebib} 

\end{document}